\documentclass[a4paper]{article}
\usepackage{amsmath,amssymb,amsfonts,amsthm,amscd}
\usepackage{multirow}
\usepackage{mathtools}
\usepackage[titletoc,title]{appendix}
\usepackage[bookmarks=true]{hyperref}
\usepackage{color}
\numberwithin{equation}{section}

\newcommand{\bea}{\begin{eqnarray}}
\newcommand{\eea}{\end{eqnarray}}
\newcommand{\be}{\begin{equation}}
\newcommand{\ee}{\end{equation}}
\newcommand{\nn}{\nonumber \\}

\begin{document}
\begin{titlepage}

\vfill \vfill \vfill
\begin{center}
{\bf\Large Dual superfield approach to supersymmetric mechanics\\ with spin variables}
\end{center}
\vspace{1.5cm}

\begin{center}
{\large\bf Stepan Sidorov}
\end{center}
\vspace{0.4cm}

\centerline{\it Bogoliubov Laboratory of Theoretical Physics, JINR, 141980 Dubna, Moscow Region, Russia}

\vspace{0.3cm}

\centerline{\tt sidorovstepan88@gmail.com}
\vspace{0.2cm}

\vspace{2cm}

\par

{\abstract
\noindent
We consider a reducible ${\cal N}=4$, $d=1$ multiplet described by a real superfield as a coupling of the mirror ${\bf (1, 4, 3)}$ and ordinary ${\bf (3, 4, 1)}$ multiplets. Employing this so-called ``long multiplet'', we construct a coupled system of dynamical and semi-dynamical multiplets. We show that the corresponding {\it on-shell} model reproduces the model of Fedoruk, Ivanov and Lechtenfeld presented in 2012. Furthermore, there is a hidden supersymmetry acting on the long multiplet that extends the full world-line supersymmetry to SU(2$|$2). In other words, the ${\cal N}=4$ long multiplet can be derived from an irreducible SU(2$|$2) multiplet.}
\vfill{}


\end{titlepage}

\section{Introduction}
Supersymmetric quantum mechanics (SQM) has been continuously developing since the supersymmetry breaking mechanism was introduced by Edward Witten \cite{Witten1,Witten2}. However, Hermann Nicolai was the first to introduce the simplest ${\cal N}=2$, $d=1$ superalgebra \cite{Nicolai} as
\bea
	\left\lbrace Q, \bar{Q}\right\rbrace = 2H, \qquad \left[H, Q\right]=0,\qquad \left[H,\bar{Q}\right]=0, \label{N2}
\eea
where the central charge generator $H$ was identified with the Hamiltonian.
He treated SQM as the simplest supersymmetric $d=1$ Lagrangian field theory in the framework of superfield approach. It's no wonder that models of SQM can be obtained by dimensional reduction from higher ($d>1$) dimensional supersymmetric field theories (see {\it e.g.} \cite{Denef}).
Moreover, ${\cal N}$-extended SQM may reveal more complicated target geometries than $d > 1$ theories, since it has a wider automorphism ($R$-symmetry) group ${\rm SO}\left({\cal N}\right)$ (see \cite{Glasses} and references therein).

Fedoruk, Ivanov and Lechtenfeld presented in \cite{FIL1} a superfield construction for the ${\cal N}=4$ superextension of the ${\rm U}(2)$-spin Calogero model based on the interaction of dynamical and semi-dynamical irreducible multiplets. Over the next couple of years, this work was followed by a subsequent study \cite{FIL2,BellKri,KriLech,BellKriSut,FIL3,IvKonSmi,KriLechSut,IvKon}
considering various types of dynamical multiplets coupled to the semi-dynamical multiplet ${\bf (4, 4, 0)}$\footnote{Irreducible $d=1$ multiplets of the ranks ${\cal N}=2,4,8$ are conveniently denoted as ${\bf (k,{\cal N},{\cal N}-k)}$ \cite{PashTopp,KuzRojTopp}, where ${\bf k}$ takes the values from ${\bf 0}$ to ${\cal N}$ and stands for the number of physical bosonic fields, the second number ${\cal N}$ is the number of fermionic fields and ${\bf {\cal N}-k}$ corresponds to the number of auxiliary bosonic fields.}. Later in \cite{FIL5}, the multiplet ${\bf (3, 4, 1)}$ was considered as semi-dynamical\footnote{Despite semi-dynamical multiplets are mostly associated with the initial papers \cite{FIL1,FIL2,FIL3} of Fedoruk, Ivanov and Lechtenfeld, the word ``semi-dynamical'' was introduced in the following papers \cite{IvKonSmi,IvKon}. Sometimes semi-dynamical multiplets are referred to as spin multiplets.}.

The difference between dynamical and semi-dynamical multiplets lies in their Lagrangian description. Dynamical multiplets correspond to Lagrangians that have kinetic terms for bosonic fields, {\it i.e.} terms with second-order time derivatives. Lagrangians describing semi-dynamical (or spin) multiplets have only first-order time derivatives of bosonic fields and are known in SQM as Wess-Zumino (or Chern-Simons) type Lagrangians. The Wess-Zumino Lagrangians for the irreducible multiplets ${\bf (4, 4, 0)}$ and ${\bf (3, 4, 1)}$ were constructed in the framework of harmonic superspace \cite{HSSd1}, but they were not considered there as independent invariants without kinetic terms. After quantisation, semi-dynamical bosonic fields become spin degrees of freedom.

The first example of ${\cal N}=8$ invariant model with dynamical and semi-dynamical fields was presented in \cite{FI2024}, but it was constructed in terms of ${\cal N}=4$ superfields. Recently, it was shown in \cite{KhaKriNer} that the model is ${\rm OSp}(8|2)$ superconformal.

A unique feature of ${\cal N}=4$ SQM is the existence of two equivalent class of multiplets that are ``mirror'' to each other \cite{BHSS}.
These two class are related by a permutation of ${\rm SU}(2)$ factors in the automorphism group ${\rm SO}(4)\sim{\rm SU}(2)_{\rm L} \times {\rm SU}(2)_{\rm R}$ of the corresponding superalgebra:
\bea
    \left\lbrace Q^{i}_{\alpha}, Q_{j}^{\beta}\right\rbrace = 2\,\delta^i_j\delta_\alpha^\beta H,\qquad \left[H,Q^{i}_{\alpha}\right]=0.\label{N4alg}
\eea
Here, Latin ($i,j=1,2$) and Greek ($\alpha,\beta = 1,2$) indices are ${\rm SU}(2)_{\rm L}$ and $ {\rm SU}(2)_{\rm R}$ doublet indices, respectively. 
Permuting them as $i,j \leftrightarrow \alpha,\beta $, one obtains the same algebra \eqref{N4alg}.
We focus our attention on the irreducible multiplets ${\bf (1, 4, 3)}$, which are described by real superfields $Y$ and $X$ satisfying
\bea
    {\rm (a)}\;\; D^{i(\alpha}D^{\beta)}_{i}Y = 0,\qquad 
    {\rm (b)}\;\; D^{(i}_{\alpha}D^{j)\alpha}X = 0.
\eea
We assume that the constraint ${\rm (a)}$ corresponds to the ordinary multiplet and the constraint ${\rm (b)}$ defines the mirror one. The permutation of ${\rm SU}(2)$ indices changes the roles of the multiplets, {\it i.e.} ${\rm (a)} \leftrightarrow {\rm (b)}$. General superfield Lagrangians of both multiplets correspond to dynamical descriptions.
By passing to on-shell Lagrangians and transformations, the equivalence of the ordinary and mirror systems was shown via duality transformations \cite{IvKriPash}.

The goal of the present work is to consider a special modification of the quadratic constraint ${\rm (b)}$ written as
\bea
D^{(i}_{\alpha}D^{j)\alpha}X_{\kappa} = 4i\kappa\,V^{ij} \quad \Longrightarrow \quad D^{(i}_{\alpha}V^{jk)}=0. \label{intro}
\eea 
The triplet $V^{ij}$ is a bosonic superfield that describes the ordinary multiplet ${\bf (3, 4, 1)}$.
The real superfield $X_{\kappa}$ describes a reducible but indecomposable multiplet with $V^{ij}$ being an irreducible subrepresentation. 
Such multiplets of ${\cal N}=2,4$ SQMs are known as ``non-minimal multiplets'' \cite{GonKhoTopp,GonIgaKhoTopp} or ``long multiplets'' \cite{CasimirEnergy,LM}. 

Superfield description of long multiplets implies a modification of standard constraints of irreducible multiplets with some parameters of coupling. In the quadratic constraint \eqref{intro} we deal with the real parameter $\kappa$.
We can give another example of ${\cal N}=4$ long multiplets that combines two chiral multiplets ${\bf (2, 4, 2)}$ \cite{LM}. The modified chiral constraint for $V$ is written as
\bea
    D^{2}_{i}V = \lambda\,D^{1}_{i}W,
\eea
where $W$ is a chiral superfield satisfying $D^{2}_{i}W=0$. Sending $\lambda \to 0$ we obtain two independent chiral superfields describing two irreducible multiplets ${\bf (2, 4, 2)}$. To describe ${\cal N}=4$ long multiplets, one can also exploit harmonic superfields \cite{HSSd1}. The constraint \eqref{intro} is harmonised as
\bea
D^{+}_{\alpha}D^{+\alpha}X_{\kappa} = 4i\kappa\,V^{++},\qquad D^{++}X_{\kappa} = 0 \quad \Longrightarrow \quad D^{+}_{\alpha}V^{++}=0,\qquad D^{++}V^{++}=0,
\eea
where $V^{++}$ is an analytic harmonic superfield. Alternatively, the formalism of biharmonic superspace \cite{BHSS} can serve as a tool for description of ${\cal N}=4$ long multiplets.

The paper is organised as follows.
Section \ref{N4long} is devoted to the ${\cal N}=4$ long multiplet described by $X_{\kappa}$ and $V^{ij}$. First, we consider the irreducible multiplets ${\bf (1, 4, 3)}$ and ${\bf (3, 4, 1)}$, separately.
We then combine them into a long multiplet using the quadratic constraint \eqref{intro} and solve it.
We construct the general superfield Lagrangian describing the interaction of dynamical and semi-dynamical fields. We compare the constructed model to the  model obtained in \cite{FIL5}, where the interaction of the two ordinary irreducible multiplets ${\bf (1, 4, 3)}$ and ${\bf (3, 4, 1)}$ was considered. Finally, we show that both on-shell models are equivalent via duality transformations \cite{IvKriPash}. In Section \ref{SU22SQM}, we discuss a relation of long multiplets to deformed SQMs, and show how the ${\cal N}=4$ long multiplet is derived from SU$(2|2)$ SQM \cite{SU22}. In Section \ref{Outlook}, we discuss problems for the future analysis.

\section{Reducible ${\cal N}=4$ long multiplet}\label{N4long}
It is convenient for us to work with the ${\rm SU}(2)_{\rm L} \times {\rm SU}(2)_{\rm R}$ covariant formulation of ${\cal N}=4$ SQM with the corresponding superalgebra \eqref{N4alg}.
The ${\cal N}=4$, $d=1$ superspace is parametrised by a time coordinate $t$ and a quartet of Grassmann coordinates $\theta^{i\alpha}$.
The coordinates transform as
\bea
    \delta \theta^{i\alpha}= \epsilon^{i\alpha}, \qquad  \delta t = -\,i\,\epsilon^{i\alpha}\theta_{i\alpha}\,,
    \qquad \overline{\left(\theta^{i\alpha}\right)}=-\,\theta_{i\alpha}\,,\qquad \overline{\left(\epsilon^{i\alpha}\right)}
    =-\,\epsilon_{i\alpha}\,.\label{SStr}
\eea
The covariant derivatives $D^{i\alpha}$ are defined as
\bea
    D^{i\alpha}=\frac{\partial}{\partial \theta_{i\alpha}}+i\,\theta^{i\alpha}\partial_t\,.\label{covD}
\eea

\subsection{Mirror multiplet (1,4,3)}
An arbitrary unconstrained real superfield contains 8 bosonic and 8 fermionic component fields in its general $\theta$-expansion. The mirror multiplet ${\bf (1,4,3)}$ is described by a real superfield $X$ satisfying a quadratic constraint 
\bea
	D^{(i}_{\alpha}D^{j)\alpha}X = 0,\qquad \overline{\left(X\right)}=X.\label{143}
\eea
The constraint kills the half of component fields, so the field content is reduced to 4 bosonic and 4 fermionic fields. The $\theta$-expansion of $X$ is given by
\bea
	&&X=x-\theta_{i\alpha}\psi^{i\alpha}+\frac{1}{2}\,\theta_{i\alpha}\theta^{i}_{\beta}A^{\alpha\beta}+\frac{i}{3}\,\theta_{i}^{\beta}\theta_{j\beta}\theta^{i}_{\alpha}\dot{\psi}^{j\alpha}-\frac{1}{12}\,\theta^{i}_{\alpha}\theta^{j\alpha}\theta_{i\beta}\theta^{\beta}_{j}\,\ddot{x},\nn
	&&\overline{\left(x\right)}=x,\qquad  \overline{\left(\psi^{i\alpha}\right)}=\psi_{i\alpha}\,,\qquad \overline{\left(A^{\alpha\beta}\right)}=-\,A_{\alpha\beta}\,,\qquad A^{\alpha\beta}=A^{\beta\alpha}.\label{X}
\eea
The component fields transform as
\bea
    \delta x = \epsilon_{i\alpha}\psi^{i\alpha},\qquad
    \delta \psi^{i\alpha} = \epsilon^i_{\beta}A^{\alpha\beta}+i\,\epsilon^{i\alpha}\dot{x},\qquad
    \delta A^{\alpha\beta} = 2i\,\epsilon^{i(\alpha}\dot{\psi}^{\beta)}_i.\label{tr143}
\eea
The general invariant action is constructed as
\bea
	S_{\bf (1,4,3)}==\int dt\,{\cal L}_{\bf (1,4,3)} =\frac{1}{2}\int dt\,d^4\theta\,f\left(X\right),\label{act143}
\eea
where $f\left(X\right)$ is an arbitrary function.
The component Lagrangian reads
\bea
    {\cal L}_{\bf (1,4,3)}=\left(\frac{\dot{x}^2}{2}+\frac{i}{2}\,\psi^{i\alpha}\dot{\psi}_{i\alpha}
    -\frac{A^{\alpha\beta}A_{\alpha\beta}}{4}\right)g\left(x\right)-\frac{1}{4}\,A^{\alpha\beta}\psi^i_{\alpha}\psi_{i\beta}\,g^{\prime}\left(x\right)
    -\frac{1}{24}\,\psi^{i}_{\alpha}\psi_{i\beta}\psi^{j\alpha}\psi^{\beta}_{j}\,g^{\prime\prime}\left(x\right),
\eea
where $g\left(x\right)=f^{\prime\prime}\left(x\right)$.
Eliminating the auxiliary field $A_{\alpha\beta}$ by its equation of motion, we find the relevant on-shell Lagrangian as
\bea
    {\cal L}^{\rm on-shell}_{\bf (1,4,3)} = \left(\frac{\dot{x}^2}{2}+\frac{i}{2}\,\psi^{i\alpha}\dot{\psi}_{i\alpha}\right)g\left(x\right)-\frac{1}{24}\,\psi^{i}_{\alpha}\psi_{i\beta}\psi^{j\alpha}\psi^{\beta}_{j}\left[g^{\prime\prime}\left(x\right)-\frac{3\,g^{\prime}\left(x\right)f^{\prime}\left(x\right)}{2\,g\left(x\right)}\right].\label{SU2SU2_Lag}
\eea
The on-shell transformations are
\bea
    \delta x = \epsilon_{i\alpha}\psi^{i\alpha},\qquad
	\delta \psi^{i\alpha} = -\,\frac{g^{\prime}\left(x\right)}{2g\left(x\right)}\,\epsilon^i_{\beta}\psi^{j\alpha}\psi_{j}^{\beta}+i\,\epsilon^{i\alpha}\dot{x}.
\eea
Let us redefine the fields as follows:
\bea
	y=f^{\prime}\left(x\right),\quad \dot{x}\left(t\right)=x^{\prime}\left(y\right)\dot{y}\left(t\right), \quad \tilde{g}\left(y\right)=x^{\prime}\left(y\right)= \frac{1}{y^{\prime}\left(x\right)}=\frac{1}{f^{\prime\prime}\left(x\right)},\quad \psi^{i\alpha}=\eta^{i\alpha}x^{\prime}\left(y\right).\label{new}
\eea
Then the Lagrangian \eqref{SU2SU2_Lag} is rewritten in terms of new fields $y$ and $\eta^{i\alpha}$ as
\bea
    {\cal L}^{\rm on-shell}_{\bf (1,4,3)} = \left(\frac{\dot{y}^2}{2}+\frac{i}{2}\,\eta^{i\alpha}\dot{\eta}_{i\alpha}\right)\tilde{g}\left(y\right)+\frac{1}{24}\,\eta^{i}_{\alpha}\eta_{i\beta}\eta^{j\alpha}\eta^{\beta}_{j}\left[\tilde{g}^{\prime\prime}\left(y\right)-\frac{3\,\tilde{g}^{\prime}\left(y\right)\tilde{g}^{\prime}\left(y\right)}{2\,\tilde{g}\left(y\right)}\right],
\eea
which is invariant under the transformations
\bea
    \delta y = \epsilon_{i\alpha}\eta^{i\alpha},\qquad
	\delta \eta^{i\alpha} = -\,\frac{\tilde{g}^{\prime}\left(y\right)}{2\tilde{g}\left(y\right)}\,\epsilon^{\alpha}_{j}\eta^{j\beta}\eta^{i}_{\beta}+i\,\epsilon^{i\alpha}\dot{y}.
\eea
The on-shell Lagrangian and transformations in the new notation coincide with those for the ordinary multiplet ${\bf (1,4,3)}$. This equivalence via the duality transformations \eqref{new} was discovered in \cite{IvKriPash}.

\subsection{Multiplet (3,4,1)}
The multiplet ${\bf (3,4,1)}$ is described by a triplet superfield $V^{ij}$ ($V^{ij}=V^{ji}$) that satisfies 
\bea
	D^{(i}_{\alpha}V^{jk)}=0,\qquad
	\overline{\left(V^{ij}\right)}=V_{ij}\,.\label{341}
\eea
The solution reads
\bea
	&&V^{ij}=v^{ij}-i\,\theta^{(i}_{\alpha}\chi^{j)\alpha}-\frac{i}{2}\,\theta^{i}_{\alpha}\theta^{j\alpha}C+i\,\theta^{\alpha}_{k}\theta^{(i}_{\alpha}\dot{v}^{j)k}-\frac{1}{3}\,\theta^{k\alpha}\theta_{k}^{\beta}\theta^{(i}_{\alpha}\dot{\chi}^{j)}_{\beta}+\frac{1}{12}\,\theta^{k}_{\alpha}\theta^{l\alpha}\theta_{k\beta}\theta^{\beta}_{l}\,\ddot{v}^{ij},\nn
	&&\overline{\left(v^{ij}\right)}=v_{ij}\,,\qquad v^{2}=\frac{1}{2}\,v^{ij}v_{ij}\,,\qquad \overline{\left(\chi^{i\alpha}\right)}
    =-\,\chi_{i\alpha}\,,\qquad \overline{\left(C\right)}=C.
\eea
The component fields transform as
\bea
    \delta v^{ij} = i\,\epsilon^{(i}_{\alpha}\chi^{j)\alpha},\qquad
    \delta \chi^{i}_{\alpha} = -\,2\,\epsilon_{j\alpha}\dot{v}^{ij}-\epsilon^{i}_{\alpha}C,\qquad
    \delta C = -\,i\,\epsilon_{k\alpha}\dot{\chi}^{k\alpha}.\label{tr341}
\eea
The Wess-Zumino action for the harmonised superfield $V^{++}$ was constructed as an analytic superpotential \cite{HSSd1}.
Without going into details, we write the Wess-Zumino Lagrangian as
\bea
    {\cal L}_{\rm WZ}=C\,{\cal U}\left(v\right)-\dot{v}^{ij}{\cal A}_{ij}\left(v\right)-\frac{i}{2}\,\chi^{i\alpha}\chi^{j}_{\alpha}\,{\cal R}_{ij}\left(v\right),\label{WZ_L}
\eea
where
\bea
    \partial^{ij}\partial_{ij}\,{\cal U}\left(v\right)=0,\qquad
    {\cal R}_{ij}\left(v\right)=\partial_{ij}\,{\cal U}\left(v\right),\qquad 
    \partial^{k}_{(i}\,{\cal A}_{j)k}\left(v\right)=\partial_{ij}\,{\cal U}\left(v\right).
\eea
The mirror and non-linear versions of this multiplet, treated as semi-dynamical, were considered in \cite{IS2021} and \cite{IS2023}, respectively.

\subsection{Long multiplet}

We couple the mirror multiplet ${\bf (1,4,3)}$ to the ordinary multiplet ${\bf (3,4,1)}$ by modifying the quadratic constraint \eqref{X} as
\bea
	D^{(i}_{\alpha}D^{j)\alpha}X_{\kappa} = 4i\kappa\,V^{ij},\qquad D^{(i}_{\alpha}V^{jk)}=0.\label{143+341}
\eea
At the same time, the superfield $V^{ij}$ satisfies the standard constraint \eqref{341}, so it describes the irreducible multiplet ${\bf (3,4,1)}$.
The new superfield $X_{\kappa}$ is written as a deformation of \eqref{X}:
\bea
	X_{\kappa}=X+i\kappa\,\theta_{i}^{\beta}\theta_{j\beta}\left(v^{ij}-\frac{2i}{3}\,\theta^{i}_{\alpha}\chi^{j\alpha}-\frac{i}{6}\,\theta^{i}_{\alpha}\theta^{j\alpha}C\right).\label{X_kappa}
\eea
The transformations \eqref{tr143} are modified as
\bea
	\delta x = \epsilon_{i\alpha}\psi^{i\alpha},\qquad
	\delta \psi^{i\alpha} = \epsilon^i_{\beta}A^{\alpha\beta}+i\,\epsilon^{i\alpha}\dot{x}-2i\kappa\,\epsilon^{\alpha}_{j}v^{ij},\qquad
	\delta A^{\alpha\beta} = 2i\,\epsilon^{i(\alpha}\dot{\psi}^{\beta)}_i+2\kappa\,\epsilon^{i(\alpha}\chi^{\beta)}_{i},\label{tr143g}
\eea
while the transformations \eqref{tr341} remain unchanged.
The new condition forces the components of $X$ to transform through the components of $V^{ij}$. 
The real parameter $\kappa$ is a coupling constant that has an inverse time dimension. In the limit $\kappa \to 0$, both multiplets become independent irreducible multiplets.

One can assume that the real superfield $X_{\kappa}$ is an unconstrained real superfield, since it has $8+8$ component fields. Indeed, the constraint \eqref{143+341} kills no degrees of freedom, but only singles out the irreducible multiplet ${\bf (3,4,1)}$ from $X_{\kappa}$\,.
From another point of view, there is a real superfield ${\cal W}$ that serves as a prepotential for the ordinary multiplet ${\bf (3,4,1)}$ \cite{IvSmi,IvKriLech}:
\bea
    D^{(i}_{\alpha}D^{j)\alpha}{\cal W} = 4i\,V^{ij}.
\eea
This definition of $V^{ij}$ leads directly to the constraint \eqref{341}.
The prepotential ${\cal W}$ is subjected to the gauge transformation ${\cal W} \rightarrow {\cal W} + D^{i(\alpha}D^{\beta)}_{i}\,\omega_{\alpha\beta}$\,.
In the Wess-Zumino gauge, only components of the multiplet ${\bf (3,4,1)}$ survives. Then the prepotential ${\cal W}$ takes the form
\bea
    {\cal W}=i\,\theta_{i}^{\beta}\theta_{j\beta}\left(v^{ij}-\frac{2i}{3}\,\theta^{i}_{\alpha}\chi^{j\alpha}-\frac{i}{6}\,\theta^{i}_{\alpha}\theta^{j\alpha}C\right),
\eea
and transforms as
\bea
    \delta {\cal W}= 2i\,\theta^{\alpha}_{i}\epsilon_{j\alpha}v^{ij}-\theta_{i\alpha}\theta^{i}_{\beta}\epsilon^{j\alpha}\chi^{\beta}_{j}-\frac{2}{3}\,\theta_{i}^{\beta}\theta_{j\beta}\theta^{i}_{\alpha}\epsilon^{\alpha}_{k}\dot{v}^{jk}.\label{Wtr}
\eea
The real superfield \eqref{X_kappa} is represented as $X_{\kappa} = X+\kappa\,{\cal W}$,
where the residual transformation \eqref{Wtr} is compensated by $\delta X = -\,\kappa\,\delta{\cal W}$. It should be noted that the constraint \eqref{143+341} is not gauge invariant in order to preserve all degrees of freedom. 

\subsection{Lagrangian and duality transformations}
The most general kinetic action of the long multiplet is written in terms the superfields $X_{\kappa}$ and $V^{ij}$. Here, we limit our consideration to 
the kinetic action for $X_{\kappa}$ only:
\bea
	S_{\rm long\;kin.}=\int dt\,{\cal L}_{\rm long\;kin.} = \frac{1}{2}\int dt\,d^4\theta\,f\left(X_{\kappa}\right).\label{L143+341}
\eea
We discard $V^{ij}$ in order to treat the multiplet ${\bf (3,4,1)}$ as semi-dynamical. 
The component Lagrangian reads
\bea
    {\cal L}_{\rm long\;kin.}&=&\left(\frac{\dot{x}^2}{2}+\frac{i}{2}\,\psi^{i\alpha}\dot{\psi}_{i\alpha}
    -\frac{A^{\alpha\beta}A_{\alpha\beta}}{4}\right)g\left(x\right)-\frac{1}{4}\,A^{\alpha\beta}\psi^i_{\alpha}\psi_{i\beta}\,g^{\prime}\left(x\right)
    -\frac{1}{24}\,\psi^{i}_{\alpha}\psi_{i\beta}\psi^{j\alpha}\psi^{\beta}_{j}\,g^{\prime\prime}\left(x\right)\nn
    &&-\,\kappa\,C\,f^{\prime}\left(x\right)-\kappa^2\,v^{ij}v_{ij}\,g\left(x\right)+\kappa\,\psi^{i\alpha}\chi_{i\alpha}\,g\left(x\right)-\frac{i}{2}\,\kappa\,v^{ij}\psi^{\alpha}_{i}\psi_{j\alpha}\,g^{\prime}\left(x\right), \label{kin}
\eea
where $g\left(x\right)=f^{\prime\prime}\left(x\right)$. We see that the Lagrangian contains no time derivatives of $v^{ij}$. This means that the bosonic field is an auxiliary field, so it can be eliminated by using its equation of motion. To avoid this elimination, we add the Wess-Zumino Lagrangian \eqref{WZ_L}:
\bea
	{\cal L}_{\rm tot.}={\cal L}_{\rm long\;kin.}+\gamma\,{\cal L}_{\rm WZ}\,.
\eea
The total Lagrangian ${\cal L}_{\rm tot.}$ describes the interaction of the dynamical and semi-dynamical multiplets:
\bea
    {\cal L}_{\rm tot.}&=&\left(\frac{\dot{x}^2}{2}+\frac{i}{2}\,\psi^{i\alpha}\dot{\psi}_{i\alpha}
    -\frac{A^{\alpha\beta}A_{\alpha\beta}}{4}+\kappa\,\psi^{i\alpha}\chi_{i\alpha}-\kappa^2 v^{ij}v_{ij}\right)g\left(x\right)+C\left[\gamma\,{\cal U}\left(v\right)-\kappa\,f^{\prime}\left(x\right)\right]\nn
    &&-\,\gamma\,\dot{v}^{ij}{\cal A}_{ij}\left(v\right)-\frac{i}{2}\,\gamma\,\chi^{i\alpha}\chi^{j}_{\alpha}\,{\cal R}_{ij}\left(v\right)-\frac{1}{4}\,A^{\alpha\beta}\psi^i_{\alpha}\psi_{i\beta}\,g^{\prime}\left(x\right)
    -\frac{i}{2}\,\kappa\,v^{ij}\psi^{\alpha}_{i}\psi_{j\alpha}\,g^{\prime}\left(x\right)\nn
    &&-\,\frac{1}{24}\,\psi^{i}_{\alpha}\psi_{i\beta}\psi^{j\alpha}\psi^{\beta}_{j}\,g^{\prime\prime}\left(x\right). \label{tot}
\eea
We eliminate the auxiliary fields $A^{\alpha\beta}$ and $\chi^{i\alpha}$ by their equations of motion and keep the auxiliary field $C$ as a Lagrange multiplier. Then the on-shell Lagrangian is written as
\bea
    {\cal L}^{\rm on-shell}_{\rm tot.}&=&\left(\frac{\dot{x}^2}{2}+\frac{i}{2}\,\psi^{i\alpha}\dot{\psi}_{i\alpha}-\kappa^2 v^{ij}v_{ij}\right)g\left(x\right)+C\left[\gamma\,{\cal U}\left(v\right)-\kappa\,f^{\prime}\left(x\right)\right]-\gamma\,\dot{v}^{ij}{\cal A}_{ij}\left(v\right)\nn
    &&-\,i\,\psi^{\alpha}_{i}\psi_{j\alpha}\left[\frac{\kappa}{2}\,v^{ij}g^{\prime}\left(x\right)+\frac{\kappa^{2}g^{2}\left(x\right){\cal R}^{ij}\left(v\right)}{\gamma^2\,{\cal R}^{kl}\left(v\right){\cal R}_{kl}\left(v\right)}\right]\nn
    &&-\,\frac{1}{24}\,\psi^{i}_{\alpha}\psi_{i\beta}\psi^{j\alpha}\psi^{\beta}_{j} \left[g^{\prime\prime}\left(x\right)-\frac{3\,g^{\prime}\left(x\right)g^{\prime}\left(x\right)}{2\,g\left(x\right)}\right].
    \label{tot-onshell}
\eea
The on-shell transformations are
\bea
    &&\delta x = \epsilon_{i\alpha}\psi^{i\alpha},\qquad
	\delta \psi^{i\alpha} = -\,\frac{g^{\prime}\left(x\right)}{2g\left(x\right)}\,\epsilon^i_{\beta}\psi^{j\alpha}\psi_{j}^{\beta}+i\,\epsilon^{i\alpha}\dot{x}-2i\kappa\,\epsilon^{\alpha}_{j}v^{ij},\nn
	&&\delta v^{ij} = -\,\frac{2\kappa g\left(x\right)}{\gamma\,{\cal R}^{kl}\left(v\right){\cal R}_{kl}\left(v\right)}\,\epsilon^{(i}_{\alpha} {\cal R}^{j)m}\left(v\right)\psi^{\alpha}_{m},\qquad
    \delta C = \epsilon_{i\alpha}\,\partial_t\left[\frac{2\kappa g\left(x\right)\psi^{\alpha}_{j}{\cal R}^{ij}\left(v\right)}{\gamma\,{\cal R}^{kl}\left(v\right){\cal R}_{kl}\left(v\right)}\right].\label{tr-onshell}
\eea
Finally, performing the duality transformations \eqref{new}, we rewrite the Lagrangian \eqref{tot-onshell} as
\bea
    {\cal L}^{\rm on-shell}_{\rm tot.}&=&\left(\frac{\dot{y}^2}{2}+\frac{i}{2}\,\eta^{i\alpha}\dot{\eta}_{i\alpha}\right)\tilde{g}\left(y\right)-\frac{\kappa^2 v^{ij}v_{ij}}{\tilde{g}\left(y\right)}+C\left[\gamma\,{\cal U}\left(v\right)-\kappa\,y\right]-\gamma\,\dot{v}^{ij}{\cal A}_{ij}\left(v\right)\nn
    &&+\,i\,\eta^{\alpha}_{i}\eta_{j\alpha}\left[\frac{\kappa\,v^{ij} \tilde{g}^{\prime}\left(y\right)}{2\tilde{g}\left(y\right)}-\frac{\kappa^{2}{\cal R}^{ij}\left(v\right)}{\gamma^2 {\cal R}^{kl}\left(v\right){\cal R}_{kl}\left(v\right)}\right]\nn
    &&+\,\frac{1}{24}\,\eta^{i}_{\alpha}\eta_{i\beta}\eta^{j\alpha}\eta^{\beta}_{j}\left[\tilde{g}^{\prime\prime}\left(y\right)-\frac{3\,\tilde{g}^{\prime}\left(y\right)\tilde{g}^{\prime}\left(y\right)}{2\,\tilde{g}\left(y\right)}\right],
\eea
and the on-shell transformations \eqref{tr-onshell} as
\bea
    &&\delta y = \epsilon_{i\alpha}\eta^{i\alpha},\qquad
	\delta \eta^{i\alpha} = -\,\frac{\tilde{g}^{\prime}\left(y\right)}{2\tilde{g}\left(y\right)}\,\epsilon^{\alpha}_{j}\eta^{j\beta}\eta^{i}_{\beta}+i\,\epsilon^{i\alpha}\dot{y}-\frac{2i\kappa}{\tilde{g}\left(y\right)}\,\epsilon^{\alpha}_{j}v^{ij},\nn
	&&\delta v^{ij} = -\,\frac{2\kappa}{\gamma\,{\cal R}^{kl}\left(v\right){\cal R}_{kl}\left(v\right)}\,\epsilon^{(i}_{\alpha} {\cal R}^{j)m}\left(v\right)\eta^{\alpha}_{m},\qquad
    \delta C = \epsilon_{i\alpha}\,\partial_t\left[\frac{2\kappa \eta^{\alpha}_{j} {\cal R}^{ij}\left(v\right)}{\gamma\,{\cal R}^{kl}\left(v\right){\cal R}_{kl}\left(v\right)}\right].
\eea
These on-shell Lagrangian and transformations coincide exactly with those written in \cite{FIL5}. Thus, the model has a dual superfield approach. 
\begin{table}[h!]
\centering
\begin{tabular}{|c|cc|}
\cline{1-3}
    & \multicolumn{1}{c|}{Approach I} & Approach II   \\ \cline{1-3}
    \multirow{2}{*}{off-shell} & \multicolumn{1}{c|}{$\left(y, \eta^{i\alpha}, A^{ij}\right)$ + $\left(v^{ij}, \chi^{i\alpha}, C\right)$} &  $\left(x, \psi^{i\alpha},A^{\alpha\beta}, \left(v^{ij},\chi^{i\alpha},C\right)\right)_{\kappa}$ \\ \cline{2-3}
    & \multicolumn{1}{c|}{${\cal L}_{\rm I} = {\cal L}_{\rm kin.} + \kappa\,{\cal L}_{\rm int.} + {\cal L}_{\rm WZ}$} & ${\cal L}_{\rm II} = {\cal L}_{\rm long\;kin.}+{\cal L}_{\rm WZ}$ \\ \cline{1-3}
\multirow{2}{*}{on-shell} & \multicolumn{1}{c|}{$y, \eta^{i\alpha}, v^{ij}, C$ } &  $x, \psi^{i\alpha},v^{ij},C$  \\ \cline{2-3}
    & \multicolumn{2}{c|}{ duality transformations $\Rightarrow$ ${\cal L}_{\rm I}\equiv {\cal L}_{\rm II}$}     \\ \cline{1-3}
\end{tabular}
\caption{The first approach I corresponds to the construction via irreducible multiplets \cite{FIL5}. The presented here approach II is based on the reducible ${\cal N}=4$ long multiplet.}\label{table}
\end{table}

We present schematically two approaches in the table \ref{table}. On the one hand, the coupling constant $\kappa$ emerges as some parameter in front of the interacting term $\kappa\,{\cal L}_{\rm int.}$\,. On the other hand, it is a parameter that combines the irreducible multiplets into the reducible long multiplet via the constraint \eqref{143+341}.

\section{Long multiplet from SU$(2|2)$ SQM}\label{SU22SQM}

A long multiplet of ${\cal N}=2$ SQM was obtained from SU$(2|1)$ SQM as a result of the decomposition of irreducible chiral multiplets into ${\cal N}=2$ multiplets \cite{CasimirEnergy,LM}. SU$(2|1)$ SQM is a deformation of ${\cal N} = 4$ SQM by a parameter $m$ \cite{DSQM,SKO}\footnote{Models of SU(2$|$1) SQM are also known as ``Weak supersymmetry'' models and were first considered at the component level in \cite{Weak,BelNer1,BelNer2,Romelsberger1,Romelsberger2}. SU(2$|$1) SQM can be obtained by dimensional reduction from the ${\cal N}=1$, $d=4$ supersymmetric field theories on $\mathbb{R}\times S^3$ \cite{rigid1,rigid2,rigid3}.}. The modified ${\cal N}=2$ superfield constraint couples irreducible chiral multiplets ${\bf (2, 2, 0)}$ and ${\bf (0, 2, 2)}$ into the long multiplet \cite{LM}:
\bea
\bar{D} \Psi = -\,\sqrt{2}\,m\,Z,\qquad \bar{D}Z=0.\label{N2long}
\eea
The superfield $\Psi$ can be considered as an unconstrained fermionic complex superfield exhibiting $4$ fermionic and $4$ bosonic components. In fact, $\bar{D}\Psi$ automatically satisfies the chiral condition since $\bar{D}^2=0$. The constraint \eqref{N2long} introduces the parameter $m$ and identifies an irreducible subrepresentation of $\Psi$ with the chiral superfield $Z$. The limit $m=0$ decouples them and the long multiplet becomes fully reducible.
By analogy with the long multiplet \eqref{N2long}, we derive below the long multiplet \eqref{143+341} from an irreducible multiplet of SU(2$|$2) SQM.

\subsection{Basics of SU(2$|$2) SQM}
Models of SU(2$|$2) SQM as deformations of ${\cal N}=8$ SQM models were studied at the superfield level in \cite{SU22}. The corresponding superalgebra $su(2|2)$ is written as a deformation of the ${\cal N}=8$, $d=1$ superalgebra\footnote{There are some differences in the definitions of the superalgebra $su(2|2)$ here and in \cite{SU22}. In order for the definitions to match, it is necessary to make the following redefinitions:
$Q^{i}_{\alpha}\rightarrow iQ^{i}_{a}$, $S^{i}_{\alpha}\rightarrow iS^{i}_{a}$, $I^{ij}\rightarrow L^{ij}$ and $F^{\alpha\beta}\rightarrow R^{ab}$.}
\bea
    &&\left\lbrace Q^{i}_{\alpha}, Q_{j}^{\beta}\right\rbrace = 2\,\delta^{i}_{j}\delta_{\alpha}^{\beta}H,\qquad 
    \left\lbrace S^{i}_{\alpha}, S_{j}^{\beta}\right\rbrace = 2\,\delta^{i}_{j}\delta_{\alpha}^{\beta}H,\nn
    &&\left\lbrace Q^{i}_{\alpha}, S_{j}^{\beta}\right\rbrace = 2\,\delta^{i}_{j}\delta_{\alpha}^{\beta}C - 2im\left(\delta^{\alpha}_{\beta} I^{i}_{j} + \delta^{i}_{j}F^{\alpha}_{\beta}\right),\nn
    &&\left[I^{ij}, I^{kl}\right] = \varepsilon^{il}I^{kj} +\varepsilon^{jk}I^{il}, \qquad \left[F^{\alpha\beta}, F^{\gamma\delta}\right] = \varepsilon^{\alpha\delta}F^{\beta\gamma} +\varepsilon^{\beta\gamma}F^{\alpha\delta},\nn
    &&\left[I^{ij}, Q^{k}_{\alpha}\right] = \frac{1}{2}\left(\varepsilon^{ik}Q^{j}_{\alpha} + \varepsilon^{jk}Q^{i}_{\alpha}\right), \qquad
    \left[F^{\alpha\beta}, Q^{\gamma}_{i}\right] = \frac{1}{2}\left(\varepsilon^{\alpha\gamma}Q^{\beta}_{i}+ \varepsilon^{\beta\gamma}Q^{\alpha}_{i}\right),\nn
    &&\left[I^{ij}, S^{k}_{\alpha}\right] = \frac{1}{2}\left(\varepsilon^{ik}S^{j}_{\alpha} + \varepsilon^{jk}S^{i}_{\alpha}\right), \qquad \left[F^{\alpha\beta}, S^{\gamma}_{i}\right] = \frac{1}{2}\left(\varepsilon^{\alpha\gamma}S^{\beta} _{i}+ \varepsilon^{\beta\gamma}S^{\alpha}_{i}\right).\label{su22alg}
\eea
The bosonic generators $I^{ij}$ and $F^{\alpha\beta}$ form the $su(2)_{\rm L}\times su(2)_{\rm R}$ subalgebra. Besides the Hamiltonian $H$, there is another central charge generator $C$.
The standard ${\cal N}=4$ superalgebra \eqref{N4alg} is a subalgebra of $su(2|2)$, and its automorphism group corresponds to the generators $I^{ij}$ and $F^{\alpha\beta}$.
 
The superspace is parametrised by a time coordinate $t$ and two quartets of Grassmann coordinates $\theta^{i\alpha}$ and $\hat{\theta}^{i\alpha}$.
The coordinates transform as
\bea
    && \delta \theta^{i\alpha} =\epsilon^{i\alpha} - 2im\,\theta^{i\beta}\theta^{j\alpha}\hat{\epsilon}_{j\beta}\,, \qquad
    \delta  \hat{\theta}^{i\alpha} =
\hat{\epsilon}^{i\alpha} + 2im \left[\hat{\theta}^{j(\beta}\theta^{\alpha)}_j\hat{\epsilon}^i_\beta +  \hat{\theta}^{(j}_\beta\theta^{i)\beta}\hat{\epsilon}^\alpha_j\right],\nn
    && \delta  t =  -\,i\,\hat{\epsilon}_{i\alpha}\hat{\theta}^{i\alpha}  - i\,\epsilon_{i\alpha}\theta^{i\alpha} 
    + \frac{2m}{3}\,\theta^{i\beta}\theta^{j\alpha}\theta_{j\beta}\hat{\epsilon}_{i\alpha}\,.\label{tr}
\eea
One can see that the $\epsilon^{i\alpha}$-transformations coincides with the ${\cal N}=4$ transformations \eqref{SStr}.
The SU(2$|$2) covariant derivatives are given by explicit expressions
\bea
    &&D^{i\alpha} = \frac{\partial}{\partial \theta_{i\alpha}} + i\left(\theta^{i\alpha}+\frac{2i}{3}\,m\,\hat{\theta}^{i\beta}\hat{\theta}^{j\alpha}\hat{\theta}_{j\beta}\right)\partial_{t} + 2\,\hat{\theta}^{i\alpha}\tilde{C} + 2im\,\hat{\theta}^{i\beta}\hat{\theta}^{j\alpha}\frac{\partial}{\partial \hat{\theta}^{j\beta}}+ 2im\left[\hat{\theta}_{j}^{\alpha}\,\tilde{I}^{ij} - \hat{\theta}_{\beta}^{i}\,\tilde{F}^{\alpha\beta}\right], \nn
    &&\nabla^{i\alpha} = \frac{\partial}{\partial \hat{\theta}_{i\alpha}} + i\,\hat{\theta}^{i\alpha}\partial_{t}\,.
\eea
They satisfy the anticommutation relations
\bea
    &&\left\lbrace D^{i\alpha}, D^{j\beta}\right\rbrace = 2i\,\varepsilon^{ij}\varepsilon^{\alpha\beta}\partial_{t}\,,\qquad
    \left\lbrace \nabla^{i\alpha}, \nabla^{j\beta}\right\rbrace = 2i\,\varepsilon^{ij}\varepsilon^{\alpha\beta}\partial_{t}\,,\nn
    &&\left\lbrace D^{i\alpha},\nabla^{j\beta}\right\rbrace = 2\,\varepsilon^{ij}\varepsilon^{\alpha\beta}\tilde{C}+2im\left(\varepsilon^{\alpha\beta} \tilde{I}^{ij} - \varepsilon^{ij}\tilde{F}^{\alpha\beta} \right). \label{CovAlg}
\eea
Here, $\tilde{I}^{ij}$ and $\tilde{F}^{\alpha\beta}$ are ``matrix'' parts of the full ${\rm SU}(2)$ generators, which act on the external ${\rm SU}(2)_{\rm L}\times {\rm SU}(2)_{\rm R}$ indices of superfields. On the covariant derivatives they act as
\bea
    &&\tilde{I}^{ij}D^{k\alpha} = -\,\frac{1}{2}\left(\varepsilon^{ik}D^{j\alpha} + \varepsilon^{jk}D^{i\alpha}\right), \qquad
    \tilde{F}^{\alpha\beta}D^{k\gamma} = -\,\frac{1}{2}\left(\varepsilon^{\alpha\gamma}D^{k\beta} + \varepsilon^{\beta\gamma}D^{k\alpha}\right),\nn
    &&\tilde{I}^{ij}\nabla^{k\alpha} = -\,\frac{1}{2}\left(\varepsilon^{ik}\nabla^{j\alpha} + \varepsilon^{jk}\nabla^{i\alpha}\right), \qquad
    \tilde{F}^{\alpha\beta}\nabla^{k\gamma} = -\,\frac{1}{2}\left(\varepsilon^{\alpha\gamma}\nabla^{k\beta} + \varepsilon^{\beta\gamma}\nabla^{k\alpha}\right).
\eea
Superfields can also have a representation with respect to the central charge $\tilde{C}$.

\subsection{Multiplet (4,8,4)}\label{484}
There are several variants of irreducible SU(2$|$2) multiplets with the field content ${\bf (4,8,4)}$\footnote{The variety of ${\cal N}=8$ multiplets was constructed in \cite{ABC}.}.
One of them is described by a pair of superfields ${\cal V}^{ij}$ and ${\cal X}$
satisfying 
\bea
    &&D^{(i}_{\alpha} {\cal V}^{jk)}=0,\qquad
    \nabla^{(i}_{\alpha} {\cal V}^{jk)}=0,\qquad \tilde{C} {\cal V}^{ij}=0,\qquad {\cal V}^{ij}={\cal V}^{ji},\qquad \overline{\left({\cal V}^{ij}\right)}={\cal V}_{ij}\,,\nn
    &&D^{i\alpha} {\cal V}^{jk}=-\,\varepsilon^{i(j}\nabla^{k)\alpha}{\cal X},\qquad  \nabla^{i\alpha} {\cal V}^{jk}=-\,\varepsilon^{i(j}D^{k)\alpha}{\cal X},\qquad  \tilde{C} {\cal X}=0,\qquad  \overline{\left({\cal X}\right)}={\cal X}.
\eea
The real superfield ${\cal X}$ is scalar, while $\tilde{I}^{ij}$ acts on the triplet ${\cal V}^{kl}$ as 
\bea
    \tilde{I}^{ij}{\cal V}^{kl} = -\,\frac{1}{2}\left(\varepsilon^{ik}{\cal V}^{jl} + \varepsilon^{jk}{\cal V}^{il}+\varepsilon^{il}{\cal V}^{jk} + \varepsilon^{jl}{\cal V}^{ik}\right).
\eea
Taking this and \eqref{CovAlg} into account, we impose on ${\cal X}$ quadratic constraints and derive that
\bea
    D^{(i}_{\alpha}D^{j)\alpha}{\cal X} = -\,4im\,{\cal V}^{ij},\qquad \nabla^{(i}_{\alpha}\nabla^{j)\alpha}{\cal X} = 4im\,{\cal V}^{ij}.
\eea
If we weaken the SU$(2|2)$ supersymmetry to the ${\cal N}=4$ supersymmetry by putting $\hat{\theta}^{i\alpha}=0$, then $D^{i\alpha}$ takes the explicit form \eqref{covD} and $\nabla^{i\alpha}$ vanishes. 
Hence, the multiplet ${\bf (4,8,4)}$ becomes the long multiplet \eqref{143+341}, where 
\bea
    X_{\kappa}={\cal X}|_{\hat{\theta}=0}\,,\qquad V^{ij}={\cal V}^{ij}|_{\hat{\theta}=0}\,,\qquad\kappa = -\,m.
\eea
Under the hidden supersymmetry $S^{i}_{\alpha}$\,, the component fields transform as 
\bea
    &&\delta x = i\,\hat{\epsilon}_{i\alpha}\chi^{i\alpha},\qquad
    \delta \psi^{i}_{\alpha} = -\,2i\,\hat{\epsilon}_{j\alpha}\dot{v}^{ij}-i\,\hat{\epsilon}^{i}_{\alpha}C,\qquad \delta A^{\alpha\beta} = -\,2\,\hat{\epsilon}^{i(\alpha}\dot{\chi}^{\beta)}_i+2i\kappa\,\hat{\epsilon}^{i(\alpha}\psi^{\beta)}_{i},\nn
    &&\delta v^{ij} = \hat{\epsilon}^{(i}_{\alpha}\psi^{j)\alpha},\qquad \delta \chi^{i\alpha} = -\,i\,\hat{\epsilon}^i_{\beta}A^{\alpha\beta}+\hat{\epsilon}^{i\alpha}\dot{x}+2\kappa\,\hat{\epsilon}^{\alpha}_{j}v^{ij},\qquad
    \delta C = -\,\hat{\epsilon}_{k\alpha}\dot{\psi}^{k\alpha}.\label{N4hiddencomp}
\eea
We can switch the roles of the original and hidden ${\cal N}=4$ supersymmetries in the superalgebra \eqref{su22alg}. This means that the long multiplet \eqref{143+341} can be defined alternatively via the covariant derivative $\nabla^{i\alpha}$ with the deformation parameter $\kappa=m$. Indeed, in the limit $\kappa=0$ the multiplet ${\bf (4,8,4)}$ decomposes into the multiplets ${\bf (1,4,3)}$ and ${\bf (3,4,1)}$ with switched fermions $\psi^{i\alpha} \leftrightarrow i\chi^{i\alpha}$.

\section{Outlook}\label{Outlook}

In this paper we considered an example of a reducible ${\cal N}=4$ long multiplet defined by the constraint \eqref{143+341}. We demonstrated a dual superfield approach to SQM with spin variables, which is outlined in the table \ref{table}. Moreover, we showed that the long multiplet can be derived from the irreducible SU$(2|2)$ multiplet ${\bf (4,8,4)}$ by weakening the SU$(2|2)$ supersymmetry to the ${\cal N}=4$ supersymmetry. 

There are several directions for further study of the long multiplet. First of all, we can consider the general ${\cal N}=4$ superfield action of $X_{\kappa}$ and $V^{ij}$:
\bea
    S_{\rm long\;kin.}=\frac{1}{2}\int dt\,d^4\theta\,f\left(X_{\kappa},V^{ij}\right).\label{GenAct}
\eea
This gives a full dynamical description for the long multiplet, where the corresponding component Lagrangian contains kinetic terms (second-order time derivatives) for the bosonic fields $x$ and $v^{ij}$. 
By requiring invariance under the hidden supersymmetry \eqref{N4hiddencomp}, SU$(2|2)$ supersymmetric Lagrangians of the multiplet ${\bf (4,8,4)}$ can be found. Furthermore, the $su(2|2)$ superalgebra \eqref{su22alg} contains also the superalgebras $su(2|1)$ and $su(1|2)$, which are mirror to each other by swapping $su(2)_{\rm L} \leftrightarrow su(2)_{\rm R}$\,. This means that the long multiplet admits two non-equivalent SU(2$|$1) and SU(1$|$2) generalisations\footnote{It is sufficient to consider the generalisation to SU(2$|$1) SQM \cite{DSQM,SKO}, since the SU(1$|$2) covariantised constraint \eqref{143+341} has its SU(2$|$1) counterpart derived by swapping SU$(2)$ indices as $i,j \leftrightarrow \alpha,\beta $. As was shown in \cite{DHSS}, there is no equivalence between SU(2$|$1) multiplets and their mirror counterparts.}. An interesting question is whether these two generalisations violate the dual approach \ref{table}.
Another challenging problem concerns the non-linear multiplet ${\bf (3,4,1)}$ satisfying \cite{IS2023}
\bea
    D^{(i}_{\alpha}V^{jk)}-\frac{1}{R}\,V^{l(i}D_{l\alpha}V^{jk)}=0.
\eea
Probably, the two approaches \ref{table} can be generalised to the non-linear case, where ${\cal L}_{\rm WZ}$ was constructed in \cite{IS2023}. For the approach II this will necessarily lead to a non-linear modification of the constraint \eqref{143+341}. A crucial point for the approach I is the construction of the interacting Lagrangian ${\cal L}_{\rm int.}$\,.

It would be interesting to describe other ${\cal N}=4$ long multiplets at the superfield level, a classification of which was given at the component level in \cite{GonKhoTopp,GonIgaKhoTopp}. Some of them may admit to the SU(2$|$1) generalisation \cite{DSQM,SKO}. Following what is shown in Subsection \ref{484}, we can try to define ${\cal N}=4$ and SU(2$|$1) long multiplets by considering SU(2$|$2) and SU(4$|$1) multiplets \cite{SU22,SU41}.

The problem of generalising long multiplets to ${\cal N}=8$ SQM certainly deserves attention. As an example, let us define a long multiplet composed of the multiplets ${\bf (2,8,6)}$ and ${\bf (6,8,2)}$ in ${\rm SU}(4)$ covariant formulation:
\bea
    \left\lbrace D^I,\bar{D}_J\right\rbrace =2i\,\delta^{I}_{J}\partial_t\,,\qquad  D^I=\frac{\partial}{\partial\theta_I} - i\,\bar{\theta}^I\partial_t\,,\qquad
    \bar{D}_J=-\,\frac{\partial}{\partial\bar{\theta}^J}+i\,\theta_J \partial_t\,.
\eea
Here, the capital indices $I,J,K,L$ ($I=1,2,3,4$) refer to the ${\rm SU}(4)$ fundamental representation. 
The reducible multiplet is described by a chiral superfield $\Phi$ with $16+16$ number of component fields:
\bea
    &&D^{I}\bar{\Phi} = 0,\qquad \bar{D}_{I}\Phi = 0,\qquad \overline{\left(\Phi\right)}=\bar{\Phi},\qquad D^I D^J\,\Phi - \frac{1}{2}\,\varepsilon_{IJKL}\,\bar{D}_K\bar{D}_L\,\bar{\Phi}  = \gamma\,V^{IJ},\nn
    &&D^{(I} V^{J)K}=0,\qquad \bar{D}_{(I}V_{J)K}=0,\qquad V^{IJ}=-\,V^{JI},\qquad\overline{\left(V^{IJ}\right)}=V_{IJ}=\frac{1}{2}\,\varepsilon_{IJKL}\,V^{KL}.\nn
\eea
Similarly to \eqref{143+341} and \eqref{N2long}, there exists a second superfield $V^{IJ}\equiv V^{[IJ]}$ that describes a subrepresentaion identified with the irreducible multiplet ${\bf (6,8,2)}$.
Probably, this ${\cal N}=8$ long multiplet can split into the ${\cal N}=4$ long multiplet \eqref{143+341} and its mirror counterpart given by
\bea
	D^{i(\alpha}D^{\beta)}_{i}Y_{\kappa} = 4i\kappa\,V^{\alpha\beta},\qquad D^{(\alpha}_{i}V^{\beta\gamma)}=0.
\eea
Another obvious thought is whether the chiral superfield $\Phi$ can serve as a prepotential for the multiplet ${\bf (6,8,2)}$.

\section*{Acknowledgements}
We thank Aital Unarov for participating in the project as part of his bachelor's thesis.

\end{document}